\documentclass[twocolumn,showpacs,preprintnumbers,amsmath,amssymb]{revtex4}

\usepackage{graphicx}
\usepackage{dcolumn}
\usepackage{bm}

\begin{document}
\title{Evidence for Dynamic Charge Stripes in the Phonons of Optimally
Doped YBCO}
\author{L. Pintschovius$^{1}$, Y. Endoh$^{2}$, D. Reznik$^{1,3}$, H. Hiraka$^{2}$,
J. M. Tranquada$^{4}$, W. Reichardt$^{1}$, H. Uchiyama$^{5}$, T.
Masui$^{5}$, S. Tajima$^{5}$}
\affiliation{$^{1}$Forschungszentrum Karlsruhe, Institut f\"ur Festk\"orperphysik, P.O.B. 3640,
D-76021 Karlsruhe, Germany\\
$^{2}$Institute for Material Research, Tohoku University,
Katahira, Aoba-ku, Sendai, 980-8577, Japan\\
$^{3}$Laboratoire L$\acute{e}$on Brillouin, C.E.A./C.N.R.S.,
F-91191 Gif-sur-Yvette, France\\
$^{4}$Physics Department, Brookhaven National Laboratory, Upton,
NY 11973, USA\\
$^{5}$Superconductivity Research Laboratory, ISTEC, Shinonome,
Koutu-ku, Tokyo 135-0062, Japan}
\date{\today}

\begin{abstract}
Inelastic neutron scattering investigations on optimally doped
YBCO revealed a very pronounced temperature dependence in the Cu-O
in-plane bond-stretching vibrations along the (010) direction: a
shift of spectral weight by at least 10 meV has been observed in a
narrow range of wave vectors halfway to the zone boundary. The
temperature evolution starts at about 200 K, well above  the
superconducting transition temperature. The displacement pattern
of the anomalous phonons indicates a dynamic one-dimensional
charge-density modulation within the planes, but excludes a direct
contribution from the Cu-O chains.
\end{abstract}

\maketitle

There is widespread belief that the electronic structure of high
temperature superconductors is characterized by an inhomogeneous
distribution of charge and spin in which the spins on the copper
atoms form antiferromagnetic stripes, separated by domain walls
carrying the charge. Although theory \cite{Zaanen,Loew} suggests
that stripe formation should be a universal property of the
cuprates, solid experimental evidence for stripes has been found
so far only for a particular compound, i.e.
La$_{1.45}$Nd$_{0.4}$Sr$_{0.15}$CuO$_4$ \cite{Tranquada}, in which
the stripes are static and apparently detrimental to
superconductivity. In superconducting compounds, the stripes are
expected to be dynamic in nature, but dynamic stripes are much
more difficult to detect and so the experimental basis in favor of
dynamic stripes is far from being solid. Inelastic neutron
scattering can be used to observe the dynamic spin structure by
the interaction between the neutron spin and the electron spin.
Magnetic fluctuations consistent with the stripe phase picture
have indeed been observed in the system La$_{2-x}$Sr$_x$CuO$_4$
\cite{Cheong,Yamada} whereas the interpretation of the results for
YBa$_2$Cu$_3$O$_{7-x}$ remains controversial \cite{Mook,Bourges}.
Inelastic neutron scattering can be used to look for fingerprints
of dynamic charge stripes as well, but in this case, evidence for
stripes is obtained only indirectly by searching for effects of
the dynamic charge structure on the atomic vibrations. That is to
say, phonons with a displacement pattern closely related to that
resulting from the inhomogeneous charge distribution are expected
to show an anomalous behavior somewhat reminiscent of the phonon
anomalies observed in one-dimensional conductors at temperatures
above the Peierls transition, i.e. as a precursor phenomenon to
charge-density-wave formation \cite{Pini2}.

In this Letter, we report observations by inelastic neutron
scattering on optimally doped YBa$_2$Cu$_3$7O$_{6.95}$ which fit
nicely to the predictions of recent theoretical work
\cite{Kaneshita,Park} as to phonon anomalies resulting from an
interaction with low-lying collective charge excitations. We show
that a pronounced downward shift of spectral weight occurs for
Cu-O in-plane bond-stretching vibrations propagating along the
b-direction within a narrow range of wave vectors on cooling below
room temperature. This result appears to be difficult to explain
other than with the development of dynamic charge stripe
correlations. We note that first results on an unusual temperature
dependence of phonons in optimally doped YBCO were already
reported in \cite{Chung}. An important step for achieving a better
understanding of these effects \cite{note} has been to study the
phonon branches of $\Delta_4$-symmetry instead of those of
$\Delta_1$-symmetry : these branches differ from each other in
that the elongations in the Cu-O bi-layer are in-phase for
$\Delta_1$-symmetry and out-of-phase for $\Delta_4$-symmetry. We
found that the behavior of the two types of bond-stretching
branches is in principle very similar. However, anticrossings of
the in-plane-polarized bond-stretching branch with c-axis-
polarized branches leads to a formidable complication in the case
of the branches of $\Delta_1$-symmetry \cite{Pini,Reznik}.

Inelastic neutron scattering measurements were performed on a
composite sample of optimally doped YBa$_2$Cu$_3$O$_{6.95}$ (T$_c$
= 93 K) consisting of three twinned single crystals with a total
volume of 1.5 cm$^3$ . Details of the sample are given elsewhere
\cite{Reznik}. The neutron experiments were carried out on the 1T
triple-axis spectrometer at the ORPHEE reactor of the Laboratoire
L$\acute{e}$on Brillouin at Saclay, France. The (220) reflection
of a Cu crystal was used to monochromatize the incident neutrons
in order to achieve high resolution. Pyrolytic graphite (PG) (002)
was used as analyzer crystal. Both crystals were horizontally and
vertically focusing. A PG filter was placed into the final beam to
suppress higher order contaminations. The measurements were
performed at various temperatures between T=10 K and 300 K. The
experiments focused on the Cu-O in-plane bond-stretching
vibrations in the (100) and in the (010) directions because it was
known from previous investigations \cite{Pini,Reznik,Reichardt}
that these branches show an anomalous dispersion at low
temperatures.

\begin{figure}
\centerline{\includegraphics[height=1.5in]{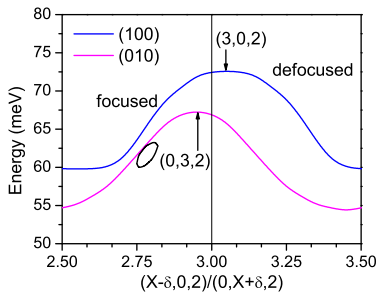}}
\caption{Dispersion relation of the out-of-phase Cu-O
bond-stretching vibrations in optimally doped YBCO as seen on a
twinned sample. The energies are plotted against the nominal wave
vector, which is calculated from an average lattice constant
a$_{av}$=(a+b)/2. The actual wavevectors for the scattering from
each of the two twin domains differ by
2$\delta$=2$\pi$/a-2$\pi$/b. The arrows show the zone center for
phonons propagating along the a or the b direction, respectively.
The ellipsoid depicts the instrumental resolution.}
\label{firstfigure}
\end{figure}

The twinning of our sample was, of course, a difficulty for the
assignment of the neutron peaks to the a or the b direction. This
problem was overcome by the technique described in refs. 9,15,
i.e. by exploiting the fact the dispersion relation for phonons
propagating along (100) is shifted in reciprocal space against
that of phonons propagating along (010) due to the difference in
lattice constants of nearly 2\% (see Fig. 1). A further strong
support for our assignment came from previous measurements on a
detwinned sample \cite{Pini,Reznik}. Unfortunately, this sample
cleaved after the first cooling cycle and could therefore not be
used for the temperature dependent measurements. As can be
inferred from Fig. 1, measurements made for momentum transfers
Q=(3,0,2)+q or Q=(3,0,2)-q, respectively, have both their pros and
cons: for Q=(3,0,2)-q, focusing leads to relatively narrow phonon
lines, but the peaks for the a* and the b* directions are rather
close in energy, whereas the opposite is true for Q=(3,0,2)+q.
Therefore, both configurations have been tried to elucidate the
behavior of the bond-stretching phonons.

A major result of our study is that strong temperature effects
were observed only for b-axis polarized phonons and only in the
vicinity of the wave vector q=(0,0.27,0). This led us to carry out
a detailed temperature study for this particular Q. We found a
pronounced shift of spectral weight from energies around 59 meV to
energies around 49 meV when lowering the temperature from 300 K to
10 K (Fig. 2a,b). The intensity loss at E$\approx$59 meV appears
to be not fully compensated by the gain at E$\approx$49 meV.
Therefore, we made further scans covering a larger energy interval
to search for the missing intensity. The data suggest that some
spectral weight is shifted down to E$\approx$37 meV (Fig. 2c) with
the caveat that it is difficult to correct for the temperature
dependence of the background  for such wide energy intervals.

Fig. 2b shows that the temperature evolution starts at
temperatures far above T$_c$ and also continues on further cooling
below T$_c$. The question as to whether there is a subtle
influence of T$_c$ on the temperature dependence of the phonon
under discussion will have to be addressed in a future
investigation using finer steps in temperature. Fig. 2b further
suggests that the phonon softening proceeds by a shift of
spectral weight and not a  peak shift, i.e. a gradual depletion
of the phonon intensity observed around 59 meV at T=300 K and a
simultaneous build-up of intensity at much lower energies. This
issue was further explored by scans at the nominal wave vector
Q=(2.7,0,2), corresponding to the same q along {\bf b*} as at
Q=(3.25,0,2) because focusing leads here to a better energy
resolution. Unfortunately, the contributions from the {\bf a*}
and the {\bf b*} directions show up at nearly the same energy at
T=200 K. On cooling, a pronounced loss in intensity is observed
which corresponds very well to that observed at Q=(3.25,0,2).
However, no peak broadening was seen at any temperature
demonstrating that all the lost intensity is shifted out of the
energy window of these scans except for a modest intensity gain
at the lowest end of the window (50 meV) (Fig. 3b). We note that
this intensity loss rapidly disappears when going to smaller q
(Fig. 3a). On the high q side, the temperature effect becomes
rapidly weaker as well and leads to a softening of the
bond-stretching phonon peak of a few meV only. At the zone
boundary, there was no discernible temperature dependence of the
bond-stretching mode (Fig. 3c).

\begin{figure}
\centerline{\includegraphics[height=3.0in]{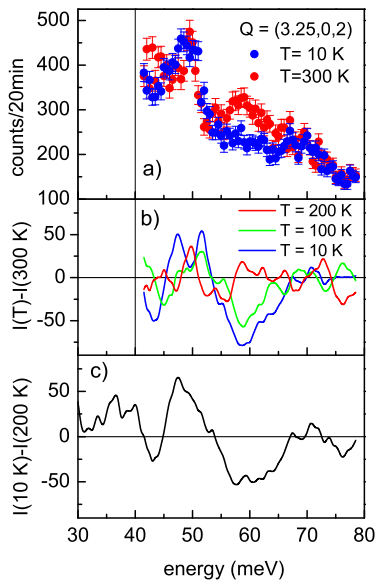}}
\caption{(a) Energy scan taken at a nominal momentum transfer of
Q=(3.25,0,2). Due to the twinning of the sample this Q corresponds
to a reduced wave vector q=0.22 along {\bf a*} or q=0.28 along
{\bf b*}, respectively. The 300 K data have been corrected for the
increase in multi-phonon scattering on raising the temperature
from T=10 K. (b) Smoothed intensity difference with respect to the
intensity measured at T=300 K. (c) Difference between the
intensities measured at T=10 K and T=200 K, respectively, as
measured in an independent run over a larger energy interval.}
\label{secondfigure}
\end{figure}

\begin{figure}
\centerline{\includegraphics[height=4.0in]{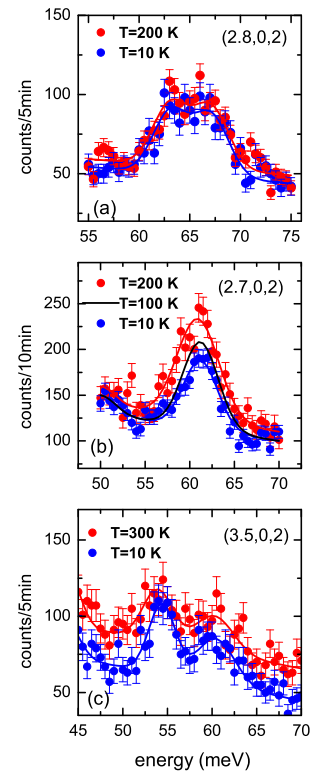}}
\caption{Energy scans taken at different temperatures. Lines
correspond to fits with one or two Gaussians. The 100 K data
points have been omitted for clarity. Due to the twinning of the
sample, the nominal wavevectors indicated in the figure correspond
to q=0.23 {\bf a*} and q=0.17 {\bf b*} (a) or q=0.33 {\bf a*} and
q=0.27 {\bf b*} (b) or q=0.47 {\bf a*} and q=0.53 ($\equiv$0.47)
{\bf b*} (c), respectively. The double peak in (a) corresponds to
the a axis mode (67 meV) and the b axis mode (63 meV),
respectively. These two peaks merge to a single one in (b) but are
again well separated in energy  in (c). The 54 meV peak in (c)
also contains contributions  of the bond-bending mode (see the
dispersion relation plotted in Fig. 4).} \label{thirdfigure}
\end{figure}

The phonon anomaly described above bears certainly some
resemblance to the Kohn anomalies observed in one-dimensional
conductors \cite{Renker}, i.e. precursor phenomena to a
charge-density-wave transition occurring at low temperatures.
Therefore, one might conjecture that it is related to flat parts
of the Fermi surface related to the oxygen chains \cite{Pickett}.
However, we emphasize that the phonon anomaly is observed in a
branch of $\Delta_4$-symmetry for which the chain O elongations
are zero. This fact excludes a trivial relationship of the anomaly
with chain O related electronic states. Having ruled out a direct
association with the Cu-O chains, we conclude that the observed
anomaly is evidence for a significant electronic anisotropy,
consistent with nematic behavior \cite{Kivelson}.

Could the phonon anomaly be the result of Fermi surface nesting
within the Cu-O planes ? A check of this possibility is provided
by recent results of density functional theory \cite{Bohnen}.
These calculations indeed predict a local frequency minimum in a
branch with bond-stretching character at {\bf
q}$\approx$(0,0.25,0); however, this branch is of
$\Delta_1$-symmetry and involves primarily chain O elongations.
This feature is not unexpected in view of the one-dimensional
nature of the Cu-O chains and may well be related to static
charge modulations detected within the Cu-O chains by scanning
tunneling microscopy (STM) \cite{Derro,Maki}. On the other hand,
the calculations do not predict a similar anomaly in a branch of
$\Delta_4$-symmetry. Rather the theoretical results are in
surprisingly good agreement with the 200 K data (Fig. 4). The
authors of that study were concerned that the high effective
electron temperature (0.2 eV) used in their calculations to
improve convergence might have suppressed the anomalous softening
at q=0.25. Therefore, additional calculations were made using a
much smaller effective electron temperature (0.02 eV). However,
this had barely any influence on the calculated phonon frequencies
\cite{Bohnenprivate}. Apparently, the anomalous low-temperature
dispersion of the in-plane bond-stretching phonons cannot be
understood within the framework of such a theory. For this
reason, it seems unlikely that the observed anomaly can be
attributed to a Fermi surface nesting effect.

A better candidate to explain the observed anomalies is the
coupling of the phonons to charge stripes, similar to what has
been proposed in recent theoretical work \cite{Kaneshita,Park}. As
indicated in Fig. 5, the displacement pattern of the anomalous
phonons is such as to favor dynamic charge accumulation on
(approximately) every fourth row of atoms along b. The idea of
charge stripe formation motivated us to search for elastic
superlattice peaks related to static charge density wave
formation, but we found none \cite{Derro}. This means that if our
interpretation of the phonon anomaly is correct, the charge
stripes in optimally doped YBCO are really dynamic in nature.

The above reasoning assumes that the charge modulation is along b,
which means that the stripes must run parallel to a, an
orientation that would be orthogonal to that implied by studies of
magnetic fluctuations in underdoped YBCO \cite{MookNature,Stock}.
On the other hand, a charge-density-wave instability (CDW) along
charge stripes has been discussed theoretically in
\cite{Kivelson}: formation of metallic stripes would make the
electronic properties of a cuprate similar to that of
one-dimensional conductors and hence could lead to the same type
of transition which is typically observed in such materials. Of
course, this argument tacitly assumes that the onset temperature
for stripe correlations is quite high which is an unresolved
issue.

\begin{figure}
\centerline{\includegraphics[height=2.0in]{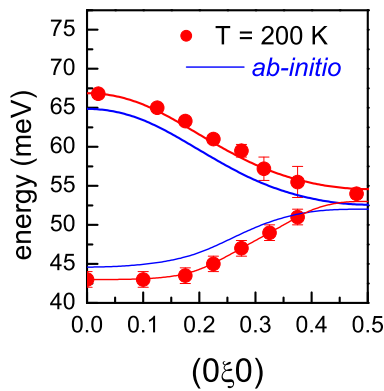}}
\caption{Comparison of the phonon dispersion of the Cu-O in-plane
bond-stretching phonons (E$>$55 meV) and of the Cu-O bond-bending
phonons (E$<$55 meV) of $\Delta_4$-symmetry as calculated from
density functional theory by \cite{Bohnen} with the experimental
data obtained at T=200 K. The red line is a guide to the eye.}
\label{fourthfigure}
\end{figure}

\begin{figure}
\centerline{\includegraphics[height=1.5in]{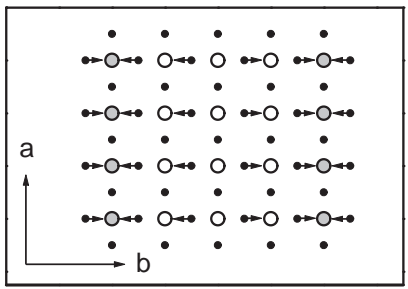}}
\caption{Displacement pattern of the longitudinal Cu-O in-plane
bond-stretching phonon with wave vector q=(0,0.25,0). Only the
elongations within one Cu-O plane are shown. The elongations in
the other layer of the bi-layer are in-phase for phonons of
$\Delta_1$-symmetry and out-of-phase for phonons of
$\Delta_4$-symmetry, respectively. Circles and full points denote
Cu atoms and O atoms, respectively. Filling of circles indicates
dynamic charge accumulation on every fourth row of atoms.}
\label{fifthfigure}
\end{figure}

Comparing the data for optimally doped YBCO (O7) with those of
underdoped O6.6 \cite{Pini} we find that the low T dispersion of
O6.6 definitely shows an anomaly similar to that observed in O7
although not quite as clearly. Re-inspection of
temperature-dependent data also revealed a significant temperature
effect which was overlooked in \cite{Pini3} because it does not
show up as a peak shift but again as a shift of spectral weight.
This indicates that the phonon anomaly reported in this paper is
not restricted to optimally doped samples.

In summary, we observed a pronounced temperature effect in the
Cu-O in-plane bond-stretching vibrations strongly indicative of
dynamic charge stripe formation in optimally doped YBCO. This
result certainly adds credence to long-held theoretical
expectations that the cuprates should form a charge-inhomogeneous
state. On the other hand, several questions remain to be answered:
Why is the effect, if present at all, much less pronounced along
the a direction ? Why is the effect not seen in optimally doped
LSCO \cite{Braden} ?  Last but not least, how does one reconcile
the magnetic response of optimally doped YBCO with the idea of a
dynamic stripe phase ? As will be discussed in a separate paper
\cite{Reznik2} antiferromagnetic spin fluctuations have indeed
been observed on the same sample, but their dispersive nature and
their apparent in-plane isotropy do not fit to the naive
expectations of the classical stripe picture \cite{Bourges2}.

This work was partially supported by the New Energy and Industrial
Technology Development Organization (NEDO) as Collaborative
Research and Development of Fundamental Technologies for
Superconductivity Applications.  Work at Brookhaven is supported
the Office of Science, U.S. Department of Energy, under Contract
No. DE-AC02-98CH10886.


\begin{references}
\bibitem{note} We note that our results and those of ref. 9 agree only on a
very qualitative level. In particular, we do not confirm the
anomalous temperature dependence of phonon intensities over the
whole Brillouin zone as suggested by Fig. 9 of Ref. 9 nor do our
data confirm the role of T$_c$ in the temperature evolution.
\bibitem{sensitivity}The sensitivity of our search was 10$^{-6}$ of the 200-Bragg peak.
\end{references}
\end{document}